\documentclass{article}
\usepackage{hiph-art}
\usepackage{psfig}
\volnumber{19} \issuenumber{1} \edyear{2004}                             
\frompage{000} \topage{000}                                              
\recrevdate{15 February 2005}                                            
\def\rightmark{ Dilepton production at SIS energies}
\def\leftmark{E.L. Bratkovskaya}

\title{Dilepton production at SIS energies }
\authors{E.L. Bratkovskaya  \\[2.812mm]
{\normalsize Institut f\"{u}r Theoretische Physik,
      Universit\"{a}t Frankfurt, 60054 Frankfurt, Germany } }

\abstract{In this short review the results of detailed studies for
dilepton production from $p+A$ and $A+A$ reactions at SIS energies are
presented. The calculations are based on a semi-classical BUU transport
model that includes the off-shell propagation of vector mesons and
evaluates  the width of the vector mesons dynamically.  Different
scenarios of in-medium modifications of vector mesons, such as
collisional broadening and dropping vector meson masses, are
investigated and the possibilities for an experimental observation of
in-medium effects in $p+A$ reactions at 1--4 GeV are discussed for a
variety of nuclear targets. }

\keyword{relativistic heavy-ion collisions, leptons}

\PACS{25.75+r; 14.60.-z;14.60.Cd}

\makeindex
\begin{document}

\maketitle

The modification of hadron properties  in nuclear matter is of
fundamental interest (cf. Refs.
\cite{BrownRho,Shakin94,Klingl96,H&L92,Asakawa93}) as QCD sum rules
\cite{H&L92,Asakawa93,Leupold} as well as QCD inspired effective
Lagrangian models
\cite{BrownRho,Shakin94,Klingl96,Herrmann,asakawa,Chanfray,Rapp,Friman,RappNPA,Peters}
predict  significant changes of the vector mesons ($\rho$, $\omega$ and
$\phi$) with the nuclear density.  A more direct evidence for the
modification of vector mesons has been obtained from the enhanced
production of lepton pairs above known sources in nucleus-nucleus
collisions at SPS energies \cite{CERES,HELIOS}.  As proposed by Li, Ko,
and Brown \cite{Li} and Ko et al. \cite{Li96}, the observed enhancement
in the invariant mass range $0.3 \leq M \leq 0.7$ GeV might be due to a
shift of the $\rho$-meson mass following Brown/Rho scaling
\cite{BrownRho} or the Hatsuda and Lee sum rule
prediction~\cite{H&L92}.  The microscopic transport studies in Refs.
\cite{Cass95C,Cass96H,Brat97,CBRep98,Ernst} for these systems support
these interpretations \cite{Li,Li96,Ko93,Ko95}.  However, also more
conventional approaches that describe a melting of the $\rho$-meson
in-medium due to the strong hadronic coupling (along the lines of
Refs.~\cite{Herrmann,asakawa,Chanfray,Rapp,Peters}) were found to be
compatible with the CERES data~\cite{Rapp,Cass95C,CBRW97}.

Dileptons have also been measured in heavy-ion collisions at the
Bevalac by the DLS collaboration \cite{DLSold,DLSnew} at incident
energies that are two orders-of-magnitude lower than that at SPS.
Although the first published spectra \cite{DLSold} based on a limited data
set are consistent with the results from transport model calculations
\cite{Xiong90,Wolf90,Gudima,BCMas96} that include $pn$ bremsstrahlung,
$\pi^0$, $\eta$ and $\Delta$ Dalitz decay and pion-pion annihilation, a
later analysis \cite{DLSnew}, including the full data set, shows a
considerable increase in the cross section, which is now more than a
factor of five above these theoretical predictions. This discrepancies
remains even after including  contributions from the decay of
$\rho$ and $\omega$ that are produced directly from nucleon-nucleon and
pion-nucleon scattering in the early reaction phase
\cite{BratRapp98,Ernst}.  With an in-medium $\rho$ spectral function, as
that used in Ref.  \cite{CBRW97} for dilepton production from heavy-ion
collisions at SPS energies, dileptons from the decay of both directly
produced $\rho$'s and pion-pion annihilation have been considered, and
a factor of two enhancement has been obtained compared to the case of
a free $\rho$-spectral function. In Ref.  \cite{BratKo99} the
alternative scenario of a dropping rho-meson mass and its influence on
the properties of the $N(1520)$ resonance has been investigated. Indeed,
an incorporation of medium effects leads to an enhancement of the rho-meson
yield, however, it was not sufficient to explain the DLS data.

As shown by the transport analysis of the Tuebingen group
\cite{Fuchs03} also another scenario for the in-medium modification,
i.e. a possible decoherence between the intermediate mesonic states in
the resonance decay, increases the dilepton yield. However, still  the
region about $M\simeq 0.3$ GeV is underestimeted whereas the yield in
the vicinity of the rho-meson peak is overestimated, especially, for
C+C collisons.  Thus, there is no consistent explaination for the DLS
data so far.

An alternative way to provide independent information about the hadron
properties in the medium is to use more elementary probes such a pions,
protons or photons as incoming particles. In such reactions the nuclear
matter is close to the ground state, i.e. at normal nuclear density,
however, in-medium effects  might be still significant to be observed
experimentally.

In Refs. \cite{Brat_pA01,Brat_KEK02}, therefore, the study of dilepton
production from heavy-ion, pion-nucleus (cf. \cite{CBRep98,Effe_piA})
and photon-nucleus reactions \cite{Effe99gam} has been extended to
proton-nucleus reactions. Within dynamically calculated width of vector
mesons the different scanarios of in-medium modifications
and there effect on the dilepton observables has been examined.
Moreover, for the first time in transport calculations the off-shell
propagation of the vector mesons -- adopted from Refs.
\cite{Cass_off1,Cass_off2} -- has be included consistently.

In this contribution a short overview of the basic futures of
off-shell dynamics for rho-meson propagation from Ref. \cite{Brat_pA01}
is presented.  This is of importance for future transport
calculations especially with respect to upcoming experimental data
from the HADES Collaboration at GSI.

\section{Description of the model}

In Ref. \cite{Brat_pA01} the analysis of dilepton production from
$p A$ collisions is performed within the BUU approach of
Refs. \cite{Effe99gam,EffePhD}.  This model is based on the
resonance concept of nucleon-nucleon and meson-nucleon interactions at
low invariant energy $\sqrt{s} \ $ \cite{TeisZP97} by adopting all
resonance parameters from the Manley analysis \cite{Manley}.

The high energy collisions -- above $\sqrt{s}$ = 2.6~GeV for
baryon-baryon collisions and $\sqrt{s}$ = 2.2~GeV for meson-baryon
collisions -- are described by  the LUND string formation and
fragmentation model FRITIOF \cite{FRITIOF}. This aspect is similar to
that used in the HSD  approach
\cite{Brat97,CBRep98,Ehehalt} and the UrQMD model \cite{Bass}.

The dilepton production within the resonance model can be schematically
presented in the following way:
\begin{eqnarray}
 BB &\to&R X   \label{chBBR} \\
 mB &\to&R X \label{chmBR} \\
      && R \to  e^+e^- X, \label{chRd} \\
      && R \to  m X, \ m\to e^+e^- X, \label{chRMd} \\
      && R \to  R^\prime X, \ R^\prime \to e^+e^- X, \label{chRprd}
\end{eqnarray}
i.e. in a first step a resonance $R$ might be produced
in baryon-baryon ($BB$) or meson-baryon ($mB$) collisions -- (\ref{chBBR}),
(\ref{chmBR}). Then this resonance can couple to  dileptons
directly -- (\ref{chRd}) (e.g., Dalitz decay of the $\Delta$ resonance:
$\Delta \to e^+e^-N$) or decays to a meson $m$ (+ baryon) -- (\ref{chRMd})
which produces dileptons via direct decays ($\rho, \omega$)
or Dalitz decays ($\pi^0, \eta, \omega$).
The resonance $R$ might also decay into another resonance $R^\prime$ --
(\ref{chRprd}) which later produces dileptons via Dalitz decay or again
via meson decays  (e.g., $D_{35}(1930)\to\Delta\rho,\ \Delta\to
e^+e^-N, \ \rho\to e^+e^-$).  Note, that in the combined model the
final particles -- which couple to dileptons -- can be produced also
via non-resonant mechanisms, i.e. 'background' at low and intermediate
energies and string decay at high energies.

The electromagnetic part of all conventional dilepton sources  --
$\pi^0, \eta, \omega$ and $\Delta$ Dalitz decay, direct decay of vector
mesons $\rho, \omega$ and $pn$ bremsstrahlung -- are treated in the
same way as described in detail in Ref.~\cite{BCM00SIS}-- where dilepton
production in $pp$ and $pd$ reactions has been studied -- and
should not be repeated here again.

\section{In-medium effects on dilepton production.}

\subsection{Collisional broadening and in-medium propagation}

In line with Refs. \cite{GKC97} the effects of collisional
broadening for the vector meson width have been implemented:
\begin{eqnarray}
\Gamma^*_V(M,|\vec p|,\rho)=\Gamma_V(M)
+ \Gamma_{coll}(M,|\vec p|,\rho) ,
\label{gammas}
\end{eqnarray}
where the collisional width is given as
\begin{eqnarray}
\Gamma_{coll}(M,|\vec p|,\rho) = \gamma \ \rho < v \ \sigma_{VN}^{tot} >.
\label{dgamma}
\end{eqnarray}
Here $v=|{\vec p}|/E, \ {\vec p}, \ E$ are the vector
meson velocity, 3-momentum and energy with respect to the target at
rest, $\gamma$ is the Lorentz factor for the boost to the rest frame
of the vector meson, $\rho$ the
nuclear density and $\sigma_{VN}^{tot}$ is the meson-nucleon total
cross section calculated within the Manley resonance model
\cite{Manley}, while $\Gamma_V(M)$ denotes the vacuum width according
to the Manley parametrization \cite{Manley} (for details see
Ref. \cite{Effe99gam}). In Eq.~(\ref{dgamma}) the brackets stand
for an average over the Fermi distribution of the nucleons.

While propagating through the nuclear medium the total width of the
vector meson $\Gamma_V^*$ (\ref{gammas}) changes dynamical and its
spectral function is modified according to the real part of the vector
meson self energy $Re \Sigma^{ret}$, as well as by the imaginary part
of the self energy ($\Gamma_V^*\simeq -Im \Sigma^{ret}/M$) following
\begin{eqnarray}
A_V(M) =  {2\over \pi} \ {M^2 \Gamma_V^*
\over (M^2-M_0^2- Re \Sigma^{ret})^2 + (M {\Gamma_V^*})^2},
\label{spfun}\end{eqnarray}
which is the in-medium form for a boson spectral function.

Since the vector mesons are produced at finite density in line with the
mass-distribution (\ref{spfun}) with $\Gamma_V^* \neq \Gamma_V$ in the
kinematical allowed mass regime, their spectral function has to change
during propagation and to merge
the vacuum spectral function when propagating out of the medium.

In Ref. \cite{Brat_pA01} the general off-shell equations of motion from
Refs.  \cite{Cass_off1,Cass_off2} have been employed which
for test particles  with momentum $\vec P_i$, energy
$\varepsilon_i$ at position $\vec X_i$ read
\begin{eqnarray}
\label{eomr} \frac{d {\vec X}_i}{dt} \! & = &
\frac{1}{2 \varepsilon_i} \: \left[ \, 2 \,
{\vec P}_i \, + \, {\vec \nabla}_{P_i} \, Re \Sigma^{ret}_{(i)} \,
+ \, \frac{ \varepsilon_i^2 - {\vec P}_i^2 - M_0^2 - Re
\Sigma^{ret}_{(i)}}{\Gamma_{(i)}} \: {\vec \nabla}_{P_i} \,
\Gamma_{(i)} \: \right],
\\[0.3cm]
\label{eomp} \frac{d {\vec P}_i}{d t} \! & = &
\frac{1}{2 \varepsilon_{i}} \: \left[ {\vec
\nabla}_{X_i} \, Re \Sigma^{ret}_i \: + \: \frac{\varepsilon_i^2 -
{\vec P}_i^2 - M_0^{2} - Re \Sigma^{ret}_{(i)}}{\Gamma_{(i)}} \:
{\vec \nabla}_{X_i} \, \Gamma_{(i)} \: \right],
\\[0.3cm]
\label{eome} \frac{d \varepsilon_i}{d t}
& = & \frac{1}{2 \varepsilon_i} \: \left[ \frac{\partial Re
\Sigma^{ret}_{(i)}}{\partial t} \: + \: \frac{\varepsilon_i^2 - {\vec
P}_i^2 - M_0^{2} - Re \Sigma^{ret}_{(i)}}{\Gamma_{(i)}} \:
\frac{\partial \Gamma_{(i)}}{\partial t} \right],
\end{eqnarray}
where the notation $F_{(i)}$ implies that the function is taken at
the coordinates of the test particle, i.e. $F_{(i)} \equiv
F(t,\vec{X}_{i}(t),\vec{P}_{i}(t),\varepsilon_{i}(t))$.
In Eqs. (\ref{eomr})-(\ref{eome}) $Re \Sigma^{ret}$ denotes the real
part of the retarded self energy while $\Gamma = -Im \Sigma^{ret} /2$
stands for the imaginary part in short-hand notation. Note, that in
(\ref{eomr})-(\ref{eome}) energy derivatives of the self energy
$\Sigma^{ret}$ have been discarded (cf. \cite{Cass_off1,Cass_off2}).

Furthermore, following Ref. \cite{Cass_off1} and using $M^{2} = P^2 -
Re \Sigma^{ret}$ as an independent variable instead of the energy
$P_0 \equiv \varepsilon$, Eq. (\ref{eome}) turns to
\begin{eqnarray}
\frac{dM_i^2}{dt} \; = \; \frac{M_i^2 -
M_0^2}{\Gamma_{(i)}} \; \frac{d \Gamma_{(i)}}{dt}
\label{eomm}
\end{eqnarray}
for the time evolution of the test-particle $i$ in the invariant
mass squared \cite{Cass_off1,Cass_off2}.

Apart from the propagation in the real potential
$\sim Re \Sigma/2\varepsilon$ the equations (\ref{eomr}) -- (\ref{eomm})
include the dynamical changes due to the imaginary part of the
self energy $Im \Sigma^{ret} \sim - M \Gamma_{V}^*$ with
$\Gamma_V^*$ from (\ref{gammas}).  It is worth to mention that the
deviation from the pole mass, i.e. $\Delta M^2 = M^2 - M_0^2$, follows
the equation
\begin{eqnarray}
{d\over dt}\Delta M^2 = {\Delta M^2\over Im \Sigma^{ret}}
\ {d\over dt} Im\Sigma^{ret},
\label{dm2}\end{eqnarray}
which expresses the fact that the off-shellness in mass is proportional
to the total width $\Gamma_V^*$.  Note, furthermore, that the
equations of motion (\ref{eomr}) -- (\ref{eomm}) conserve the particle
energy $\varepsilon$ if the self energy $\Sigma^{ret}$ does not depend
on time explicitly (cf. Refs.  \cite{Cass_off1,Cass_off2}), which is
approximately the case for $p+A$ reactions.

\subsection{'Dropping' vector meson mass}

In order to explore the observable consequences of vector meson mass
shifts at finite nuclear density the in-medium vector meson masses are
modeled according to the Hatsuda and Lee \cite{H&L92} or Brown/Rho
scaling \cite{BrownRho} as
\begin{eqnarray}
\label{Brown}
M^* = M_0 \left(1 - \alpha {\rho ({\vec r}) \over \rho_0}\right),
\end{eqnarray}
where $\rho ({\vec r})$ is the nuclear density at the resonance decay,
$\rho_0 = 0.16 \ {\rm fm}^{-3}$ and $\alpha \simeq 0.18$ for the $\rho$ and
$\omega$. The choice (\ref{Brown}) corresponds to
\begin{equation}
Re \Sigma^{ret} = M_0^2 \left( \left(\alpha \frac{\rho}{\rho_0}\right)^2
- 2 \alpha \frac{\rho}{\rho_0}\right)
\end{equation}
in (\ref{eomr}) -- (\ref{eomm}),
which is dominated by the attractive linear term in $\rho/\rho_0$
at nuclear matter density $\rho_0$.

The in-medium vector meson masses $M^*$ (\ref{Brown}) in principle have
to be taken into account in the production part as well as for
absorption reactions and for propagation. This is implemented for the
low energy reactions with nucleon resonances. Note, however, that the
vector mesons produced by the FRITIOF model -- as implemented in the
transport approach \cite{Effe99gam} -- have masses according to the
free spectral function. This approximation might not be severe since
the vector mesons from string decay at high energy have high momenta
with respect to the target nucleus where pole-mass shifts are expected
to be small \cite{Peters,Kondr_rho}.  Furthermore, the $N\rho$-width of
the baryonic resonances at finite density \cite{Effe99gam} has not been
modified.  Such modifications are out of the scope of the present
model.


\section{Dilepton spectra from $p + A$ collisions from 1--4 GeV}

\begin{figure}[t]
\phantom{a}\vspace*{5mm}
\centerline{\psfig{figure=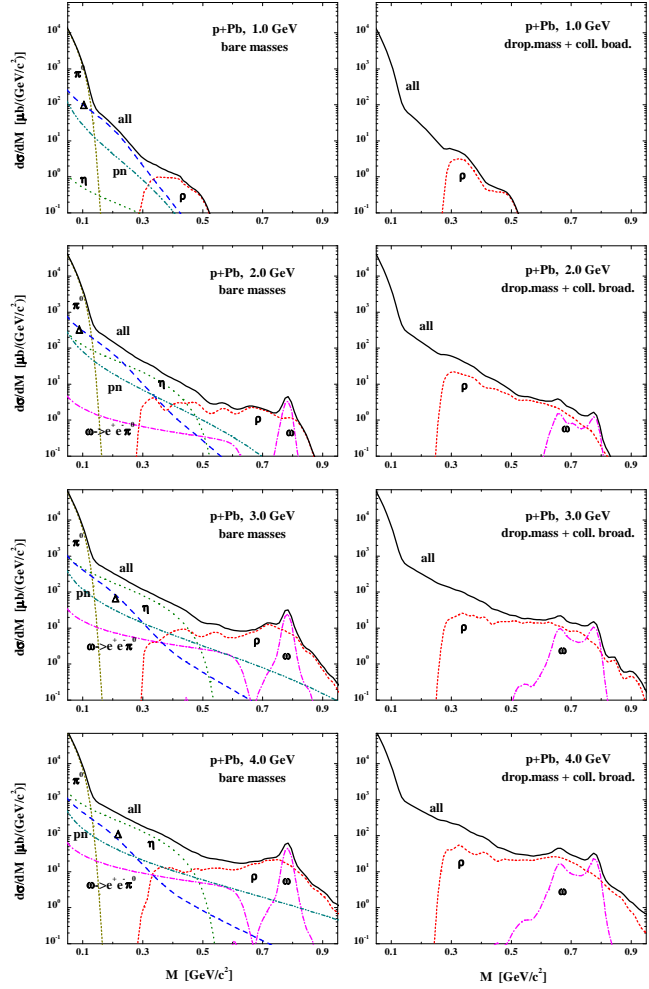,width=8.5cm}}
\caption{
The calculated dilepton invariant mass spectra $d\sigma/dM$ for $p + Pb$
collisions from 1.0 -- 4 GeV (including an experimental mass
resolution of 10 MeV)
without in-medium modifications (bare masses) -- left part,
and applying the collisional broadening + dropping masses scenario
-- right part. }
\label{Fig2pA}
\end{figure}

In Fig.~\ref{Fig2pA} the calculated dilepton invariant mass spectra
$d\sigma/dM$ are presented for $p + Pb$ collisions from
1.0 -- 4 GeV (including an experimental mass resolution
$\Delta M$ = 10 MeV) without in-medium modifications (bare masses) --
left part, and applying the collisional broadening + dropping mass
scenario -- right part.
The dominant contribution at low $M$ ($> m_{\pi^0}$) is
the $\eta$ Dalitz decay, however, for $M > 0.4$ GeV the dileptons stem
basically all from direct vector meson decays ($\rho$ and $\omega$).

\begin{figure}[!]
\phantom{a}\vspace*{5mm}
\centerline{\psfig{figure=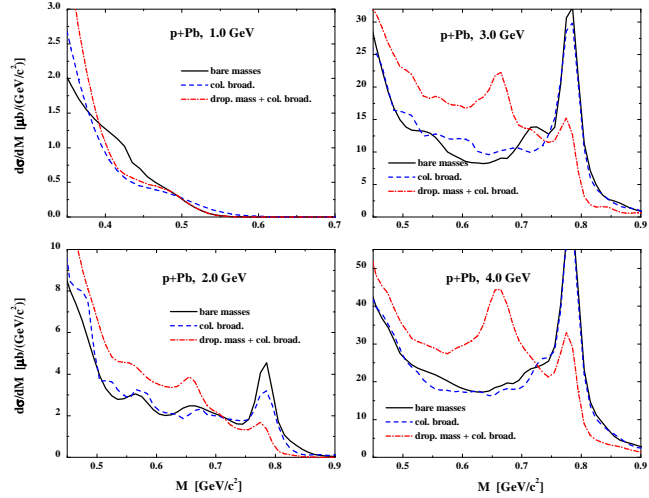,width=8.5cm}}
\caption{
The comparison of different in-medium modification scenarios, i.e.
collisional broadening (dashed lines) and collisional broadening +
dropping vector meson masses (dash-dotted lines), with respect to the
bare mass case (solid lines) on a linear scale for $p + Pb$ from 1--4 GeV.}
\label{Fig7pA}
\end{figure}

 In order to see the differences between the results from the left and
right panels of Fig. \ref{Fig2pA},  a comparison of the different
in-medium modification scenarios is shown in Fig. \ref{Fig7pA}, i.e.
collisional broadening (dashed lines) and collisional broadening +
dropping vector meson masses (dash-dotted lines), with respect to the
bare mass case (solid lines) on a linear scale for $p + Pb$ from 1--4 GeV.

Whereas collisional broadening of the $\rho$ spectral function again
gives no clear signal within the numerical accuracy achieved the
'dropping mass' scenario leads to a pronounced modification of the
spectral shape. A strong reduction of the dilepton yield in the vector
meson pole mass region around 0.77 GeV is observed since most of
the $\rho$'s and $\omega$'s now decay in the medium approximately at
density $\rho_0$. This leads to a pronounced peak around $M \approx
0.65$ GeV, which can be attributed to the in-medium $\omega$ decay
since the $\rho$ spectral strength is distributed over a wide low mass
regime. The situation is very reminiscent of dilepton spectra from $\pi
+ A$ and $\gamma + A$ reactions in Refs.
\cite{CBRep98,Effe_piA,Effe99gam}.  Especially when comparing dilepton
spectra from $C$ and $Pb$ targets, it should be experimentally possible
to distinguish an in-medium mass shift of the $\omega$ meson by taking
the ratio of both spectra.

\section{Summary}

In this contribution a short review of the detailed study in Ref.
\cite{Brat_pA01} on dilepton production  has been presented within the
framework of the coupled-channel BUU model employing a full off-shell
propagation of the vector mesons in line with Refs.
\cite{Cass_off1,Cass_off2}.  Different scenarios of in-medium
modifications of vector mesons, such as collisional broadening and
dropping vector meson masses, have been investigated and the
possibilities for an experimental observation of in-medium effects in
$p+A$ reactions has been discussed.

Dilepton spectra from $p+A$ reactions will be measured by the
HADES Collaboration at GSI Darmshtadt with high mass resolution and
good accuracy.  In this respect predictions for the dilepton invariant
mass spectra, transverse momentum and rapidity distributions for $p + A$
collisions from  1 to 4 GeV have been made in Ref. \cite{Brat_pA01}
employing different in-medium scenarios. It has been  found that the
collisional broadening + 'dropping mass' scenario leads to an
enhancement of  the dilepton yield in the range $0.5 \leq M \leq 0.75$
GeV and to a reduction of the  $\omega$-peak, which is most pronounced
for heavy systems  (up to a factor 2 for $p + Pb$ at 3--4 GeV).


\vfill\eject
\end{document}